\documentclass[aps,prl,twocolumn,superscriptaddress]{revtex4}
\usepackage{graphicx}
\usepackage{amssymb}
\usepackage{amsmath}
\usepackage{bm}

\begin{document}

\title[]{Observation of von K\'arm\'an Vortex Street in an Atomic Superfluid Gas}

\author{Woo Jin Kwon}
\affiliation{Center for Correlated Electron Systems, Institute for Basic Science, Seoul 08826, Korea}

\author{Joon Hyun Kim}
\affiliation{Center for Correlated Electron Systems, Institute for Basic Science, Seoul 08826, Korea}
\affiliation{Department of Physics and Astronomy, and Institute of Applied Physics, Seoul National University, Seoul 08826, Korea}

\author{Sang Won Seo}
\affiliation{Department of Physics and Astronomy, and Institute of Applied Physics, Seoul National University, Seoul 08826, Korea}

\author{Y. Shin}\email{yishin@snu.ac.kr}
\affiliation{Center for Correlated Electron Systems, Institute for Basic Science, Seoul 08826, Korea}
\affiliation{Department of Physics and Astronomy, and Institute of Applied Physics, Seoul National University, Seoul 08826, Korea}

\begin{abstract}
We report on the experimental observation of vortex cluster shedding from a moving obstacle in an oblate atomic Bose-Einstein condensate. At low obstacle velocities $v$ above a critical value, vortex clusters consisting of two like-sign vortices are generated to form a regular configuration like a von K\'arm\'an street, and as $v$ is increased, the shedding pattern becomes irregular with many different kinds of vortex clusters. In particular, we observe that the Stouhal number associated with the shedding frequency exhibits saturation behavior with increasing $v$. The regular-to-turbulent transition of the vortex cluster shedding reveals remarkable similarities between a superfluid and a classical viscous fluid. Our work opens a new direction for experimental investigations of the superfluid Reynolds number characterizing universal superfluid hydrodynamics.
\end{abstract}


\maketitle
The wake behind a moving obstacle is a classic subject considered in fluid dynamics. Various flow regimes are classified by the dimensionless Reynolds number $\mathrm{Re}= v D/\nu$, where $v$ is the obstacle velocity, $D$ is the lateral dimension of the obstacle, and $\nu$ is the fluid viscosity~\cite{williamson96}. At low $\mathrm{Re}<50$, a laminar or steady flow is formed, and as Re is increased, periodic shedding of vortices with alternating circulation occurs, which is known as a von K\'arm\'an vortex street. The vortex shedding frequency $f$ gives the Stouhal number $\mathrm{St}= f D/v$, which is a dimensionless quantity that is a universal function of Re.  With further increasing $\mathrm{Re}>10^5$, the wake dynamics becomes unstable and turbulent flow develops. The transition from laminar to turbulent flow represents a universal characteristic of classical fluid dynamics.

An interesting situation arises when a fluid has zero viscosity; i.e., it becomes a superfluid. The Reynolds number cannot be defined and furthermore, in contrast to classical fluids, the superfluid carries vorticity in the form of phase defects with quantized circulation. Would the superfluid show universal behavior in the wake response to a moving obstacle, and can we define a proper Reynolds number $\mathrm{Re}_s$ characterizing it~\cite{barenghi,volovik1,volovik2,bradley}? It has been clearly demonstrated that a superfluid becomes dissipative via quantum vortex emission when the obstacle velocity exceeds a critical velocity $v_c$~\cite{frisch,jackson1,winiecki2,inouye,neely,kwon1,kwon2,kwon3}. Since turbulent flow would be generated by strong perturbations of the obstacle at significantly high $v$, the key issue is whether regular vortex shedding like the von K\'arm\'an street occurs in an intermediate $v$ regime. 

\begin{figure}[b] 
\includegraphics[width=7.3cm]{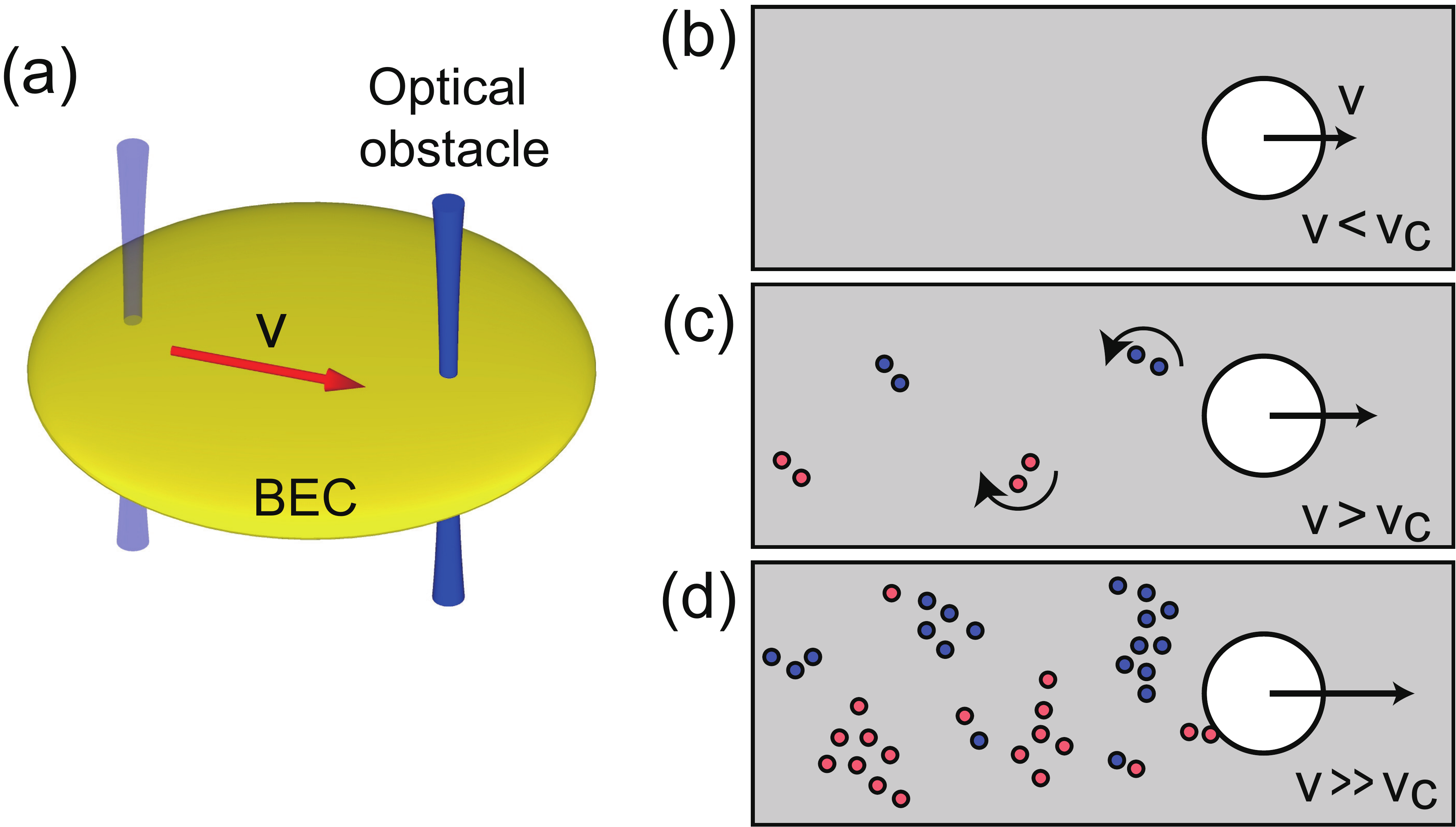}
\caption{(a) Schematic of the experiment. An impenetrable obstacle, formed by focusing a repulsive Gaussian laser beam, moves at velocity $v$ in a highly oblate Bose-Einstein condensate (BEC). Evolution of vortex shedding: (b) no excitations for $v<v_c$, (c) von K\'arm\'an street of clusters of two like-sign quantum vortices for small $v> v_c$~\cite{sasaki,stagg2,reeves1}, and (d) turbulent shedding of diversely clustered vortices for $v\gg v_c$. Red and blue circles represent vortices with clockwise and counterclockwise circulations, respectively.}
\label{scheme}
\end{figure}

Recent numerical studies of two-dimensional vortex shedding dynamics in atomic Bose-Einstein condensates (BECs) presented affirmative answers to the question~\cite{sasaki,stagg2,reeves1}. In a narrow range of $v$ above $v_c$, vortex-antivortex pairs or clusters of two like-sign vortices with alternating circulation are periodically nucleated from the obstacle [Fig.~\ref{scheme}(c)] and for high $v$, a transition to turbulence develops with irregular emission of many different kinds of large vortex clusters [Fig.~\ref{scheme}(d)]. It was noted that regular vortex shedding is stable only with clusters consisting of two like-sign vortices~\cite{sasaki,stagg2,reeves1}, and this is referred to as the quantum version of the von K\'arm\'an vortex street in a superfluid. Furthermore, Reeves \textit{et al.}~\cite{reeves1} observed for large Gaussian obstacles that St exhibits a universal relation to a superfluid Reynolds number defined as $\mathrm{Re}_s\equiv (v-v_c)D/(\hbar/m)$, where $\hbar$ is the Planck constant divided by $2\pi$ and $m$ is the particle mass, and a sudden onset of turbulence at $\mathrm{Re}_s\approx 0.7$. The proposed $\mathrm{Re}_s$ was applied in the data analysis of turbulence experiments with superfluid helium~\cite{schoepe15}.

\begin{figure*} 
\includegraphics[width=15.0cm]{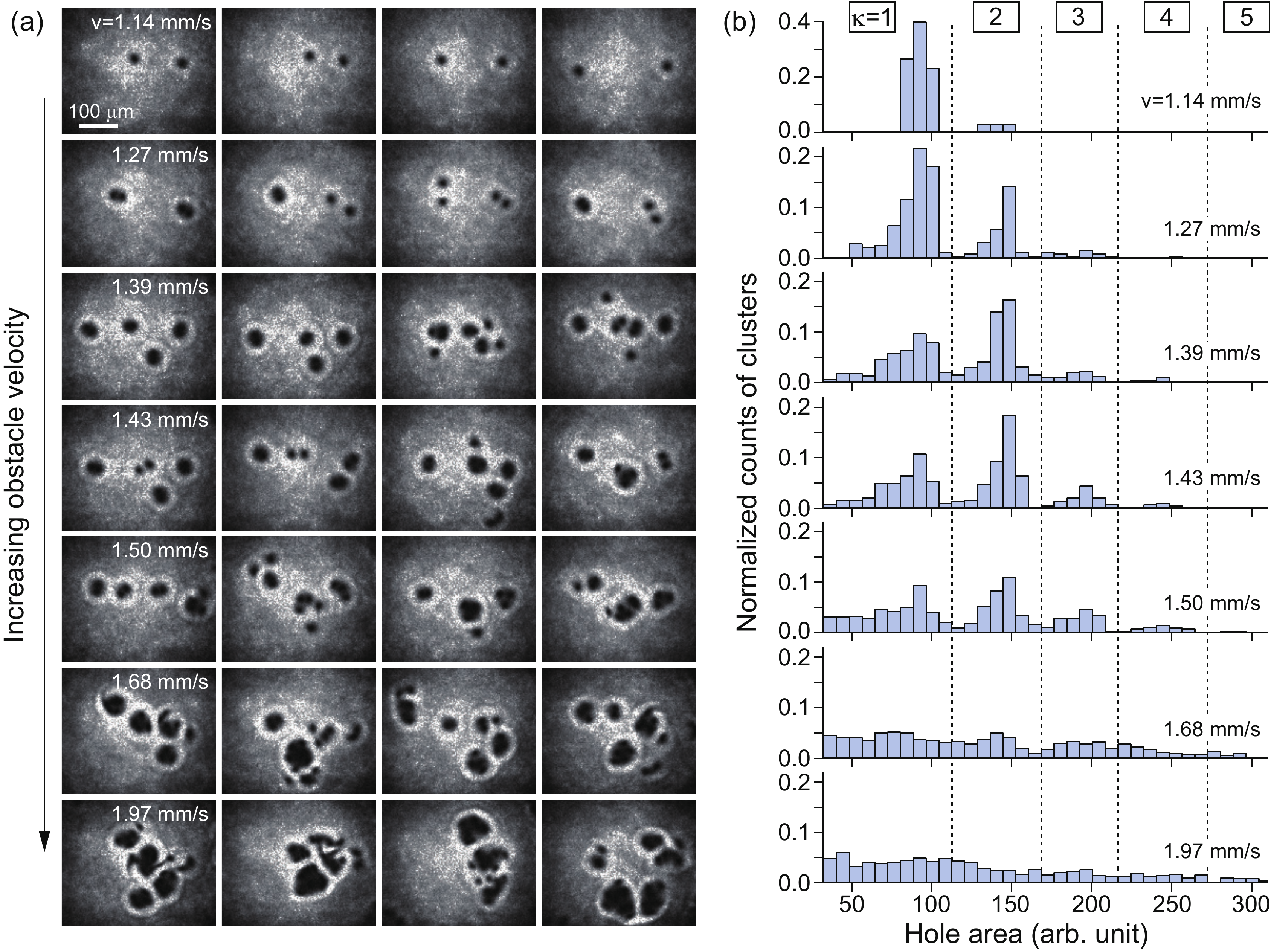}
\caption{Quantum vortex shedding from a moving optical obstacle in a highly oblate BEC. (a) Images of BECs for various obstacle velocities $v$, taken after 36~ms time of flight~\cite{tof}. Because of vortex core expansion, a vortex cluster appears with a large density-depleted hole, whose area depends on the cluster charge $\kappa$, i.e., the number of vortices in the cluster. (b) Normalized histograms of the cluster hole area. Each histogram was obtained from over 100 image data as in (a). The dashed lines indicate the transition positions for the charge number $\kappa$, which are determined from the multiple peak structure of the histograms.}
\label{image}
\end{figure*}

In this Letter, we present an experimental study of quantum vortex shedding from a moving Gaussian obstacle in a highly oblate BEC. By means of spatially large BEC samples and long-distance obstacle motion control, we examine the evolution of the vortex shedding pattern as a function of the obstacle velocity $v$. We observe regular shedding of vortex clusters each consisting of two like-sign vortices and a turbulence transition via diversifying the cluster types. Furthermore, we observe the saturation of the Stouhal number with increasing $v$, which is qualitatively consistent with the numerical results in Ref.~\cite{reeves1}. Our results demonstrate remarkable similarities between a superfluid and a classical viscous fluid in the wake response to a moving obstacle.

Our vortex shedding experiment is performed with the apparatus described in Refs.~\cite{kwon1,kwon2,kwon3}. We prepare a highly oblate BEC of $^{23}$Na atoms in a harmonic trapping potential which is generated by combining a pancake-shaped optical dipole trap and a magnetic quadruple trap. The radial and axial trapping frequencies are $\omega_{r,z}$ = $2 \pi \times (11.1, 400)$ Hz. The atom number of the condensate is $N_{0} = 5.6(4) \times 10^{6}$ and the radial Thomas-Fermi radius of the condensate is $R=\sqrt{2\mu/m\omega_r^2}\approx 105~\mu$m, where $\mu$ is the condensate chemical potential. At peak atomic density, the healing length is $\xi=\hbar/\sqrt{2m\mu} \approx 0.38~\mu$m and the speed of sound is $c_s=\sqrt{\mu/m}\approx 5.1$~mm/s. The condensate fraction of the sample is over 80$\%$. 

An optical obstacle is formed by focusing a repulsive Gaussian laser beam to the condensate [Fig.~1(a)]. The $1/e^2$ waist of the laser beam is $\sigma=10.3(11)~\mu$m $\approx 27\xi$ and its potential height is $V_0/\mu \approx 1.8$, giving the obstacle diameter $D=\sigma \sqrt{2\ln(V_{0}/\mu)}\approx 29\xi$. The obstacle position is controlled by steering the laser beam with a piezo-driven mirror. Initially, we place the obstacle at 62~$\mu$m left from the condensate center, and translate it linearly across the center region by a distance $L=114~\mu$m at a constant speed $v$ [Fig.~\ref{scheme}(a)]. After the obstacle sweeping, we turn off the laser beam linearly within 20~ms, and take an absorption image of the condensate after 36~ms time of flight~\cite{tof}. With this experimental protocol, the critical velocity for vortex shedding was measured to be $v_c = 1.11(5)~$mm/s. Note that the local condensate density varies by $\approx 35\%$ along the obstacle trajectory. At the initial and final positions, the local speed of sound and the obstacle diameter are 20\% smaller and 30\% larger than those at the center, respectively.

Figure 2(a) displays images of condensates for various obstacle velocities $v>v_c$ [Fig.~\ref{image}(a)]. In the imaging, a vortex cluster appears as a large density-depleted hole because during the time of flight vortex cores expand and would merge when they are closely located~\cite{tof}. Thus, the cluster charge $\kappa$, i.e., the vortex number of a cluster, can be inferred from the hole area. If some of vortices in a cluster have different circulation, it would be indicated by local bending of the hole shape~\cite{kwon2,kwon3}. For example, we see that some of the big clusters for high velocities $v\geq1.68~$mm/s show sharp bending tails in Fig.~2(a). When deciphering the vortex configuration from an image, it is helpful to recall that the total vortex numbers for both circulations should be the same because of angular momentum conservation.

A visual examination of images in Fig.~2(a) shows the following features of the vortex shedding. (i) Emission of like-sign vortex clusters is very likely with the obstacle for $V_0/\mu>1$. This is in contrast to the case with a penetrable obstacle ($V_0/\mu<1$), where periodic shedding of vortex dipoles was observed~\cite{kwon2,reeves2}. (ii) There is a certain low-$v$ range where $\kappa=2$ vortex clusters are dominantly generated. In particular, we observed frequent appearance of somewhat periodic shedding patterns consisting of four $\kappa=2$ clusters as shown in the first and second left images for $v=1.39$~mm/s. In terms of cluster charge regularity, this is consistent with the expected $\kappa=2$ von K\'arm\'an vortex street [Fig.~1(c)]~\cite{sasaki,stagg2,reeves1}. (iii) As $v$ is further increased, the vortex shedding pattern becomes irregular with many different larger clusters, signaling a transition to turbulence. Additional image data for various velocities are provided in the Supplemental Material~\cite{supple}.

In our experiment, the occurrence probability of the four $\kappa=2$ cluster shedding pattern was maximally about 10~\% at $v=1.39$~mm/s. Such a low probability can be attributed to the stochasticity of the incipient vortex shedding process. The vortex configuration of the four $\kappa=2$ clusters was quite reproducible~\cite{supple}, where three and one clusters are in the upper and lower regions with respect to the horizontal obstacle trajectory, respectively, which is different from the typical zigzag pattern of von K\'arm\'an street. This can be explained by the precession motion of vortices after shed from the obstacle, which is caused by the inhomogeneous density distribution of the trapped condensate~\cite{fetter}. The precession effect is most significant for the first vortex cluster. It is initially emitted at the lower side of the moving obstacle with clockwise circulation and then, it precesses to the upper region. When the first cluster has higher $\kappa$, as shown in the first image for $v=1.68$~mm/s, it moves further upward because of faster precession. The circulation direction of the first cluster seems to be deterministic due to possible asymmetry of the obstacle shape~\cite{sasaki,saito}.

To quantitatively characterize the evolution of the vortex shedding behavior, we analyze the cluster charge distribution as a function of $v$. We developed an image analysis method for quantifying the area of each density-depleted hole in an image, where the absorption image is transformed into a binary image and a particle analysis is applied~\cite{kwon3, supple}. The histogram of the hole area shows a clear multiple peak structure [Fig.~2(b)], facilitating determining the quantized charge number $\kappa$ for vortex clusters. The peak structure becomes smooth for $v>1.6$~mm/s because the vortex cluster structure is diversified and complicated by emitting more vortices.

\begin{figure}[b]
\includegraphics[width=6.7cm]{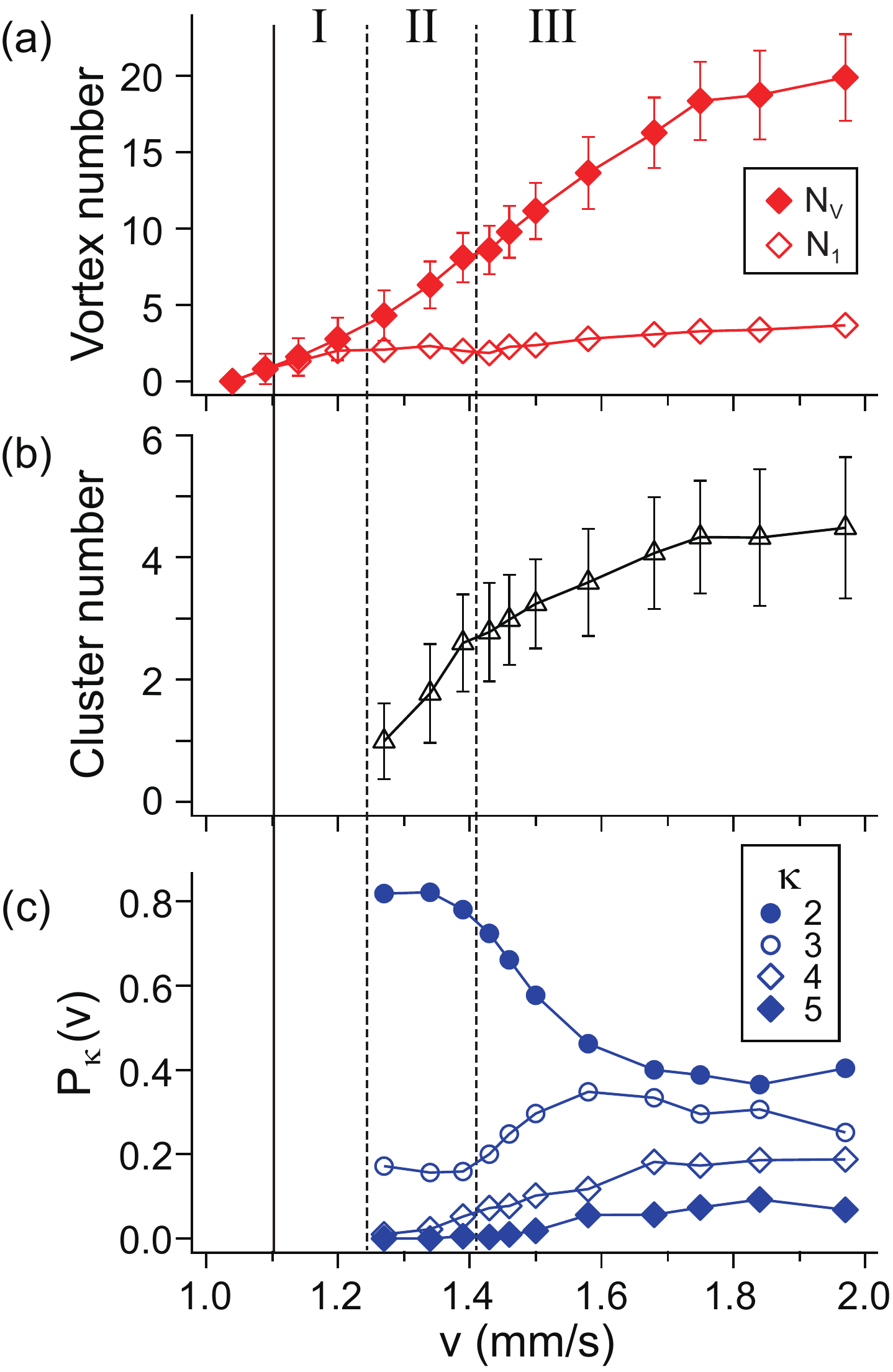}
\caption{Regular-to-turbulent transition of the quantum vortex shedding. (a) Total vortex number $N_v$ (solid red diamonds), individual ($\kappa=1$) vortex number $N_1$ (open red diamonds), (b) the number of emitted vortex clusters, $N_c$, and (c) the fractional populations $P_{\kappa}$ of charge-$\kappa$ clusters as functions of $v$. The vertical solid line denotes the critical velocity $v_c$ and the two dashed lines mark the transitions from (I) an individual vortex shedding regime to (II) a $\kappa$=2 cluster shedding regime and (III) an irregular shedding regime. The error bars indicate the standard deviation of measurements.}
\label{last}
\end{figure} 

Figure 3 displays various characteristics obtained from the cluster charge distribution. The total vortex number is estimated by $N_v=\sum_\kappa \kappa N_\kappa$, where $N_{\kappa}$ is the average number of charge-$\kappa$ clusters. The total cluster number is given by $N_c=\sum_{\kappa\geq 2} N_\kappa$, excluding individual vortices with $\kappa=1$, and the fractional population of charge-$\kappa$ clusters is $P_\kappa=N_\kappa/N_c$. We observe three velocity regimes for vortex shedding: (I) an individual vortex shedding regime with $N_v=N_1$ just above the critical velocity $v_c$, (II) a $\kappa$=2 cluster regime with $P_2\approx 0.8$ and saturated $N_1$, and (III) an irregular shedding regime where many different types of clusters are populated.

The transition to the irregular shedding regime appears pronounced with a rapid decrease of $P_2$ for $v >1.4~$mm/s. In the numerical study by Reeves {\it et~al.}~\cite{reeves1}, a similar, abrupt spreading in $P_\kappa$ was observed in the transition from the stable $\kappa=2$ cluster regime to turbulence. Furthermore, they showed that for a large obstacle, the transition occurs at the superfluid Reynolds number Re$_s\equiv {(v-v_c)D}/{(\hbar/m)}\approx 0.7$, irrespective of $D$. Our observed value of $v=1.4$~mm/s gives Re$_s\approx 1.2$ with $v_c=1.1$~mm/s. Its direct comparison to the numerical prediction is limited due to the density inhomogeneity of the trapped BEC. 

The Stouhal number St is another characteristic quantity in vortex shedding dynamics. For a cylindrical obstacle in a classical fluid, $\mathrm{St}\approx 0.2$ over a wide range of Re from $10^2$ to $10^5$. Noting that $\mathrm{St}=D/\lambda$, where $\lambda=v/f$ is the periodic spacing of vortex clusters, in our experiment the Stouhal number can be estimated approximately as $\mathrm{St}\approx (D/2L)N_c$ when $N_c/2$ cycles of cluster emission proceed over the distance $L$. In Fig.~3(b), we observe the saturating behavior of $N_c$ with increasing $v$, which is qualitatively consistent with the numerical results of Ref.~\cite{reeves1} that St increases and approaches to $\mathrm{St}_\infty\approx 0.14$ with increasing Re$_s$. Despite a large uncertainty due to the small value of $N_c$ as well as the aforementioned inhomogeneous density effect, the saturated value of $N_c\approx 4$ suggests $\mathrm{St}_\infty\sim 0.2$. We note that when we turn off the laser beam, the vortices residing in the density-vanishing region of the obstacle would be forcibly released~\cite{supple}, which can result in overestimation of St.

\begin{figure}
\includegraphics[width=5.8cm]{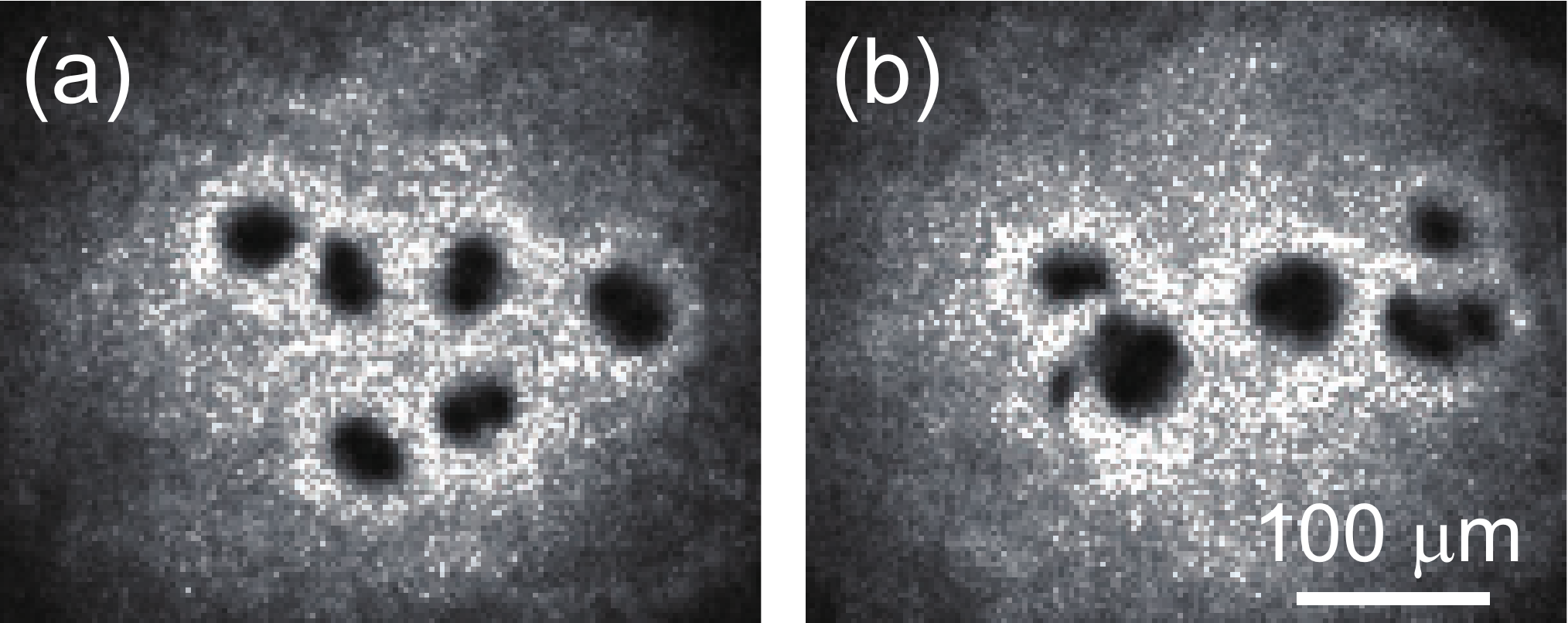}
\caption{Vortex shedding from a smaller obstacle with $\sigma/\xi\approx 18$ and $V_0/\mu \approx 3.2$. (a) Regular shedding pattern comprised of six $\kappa=2$ vortex clusters was observed at $v=1.62$~mm/s. Its appearance frequency was lower than $5\%$. (b) Typical irregular shedding pattern at the same experimental condition. The critical velocity was measured to be $v_c=1.2$~mm/s for the obstacle.}
\label{FigS1}
\end{figure}

Finally, we want to discuss the experimental requirements for stable observation of the von K\'arm\'an vortex street in a trapped BEC. First, it is necessary to have a long obstacle translation distance $L$, i.e., $L\gg D/\mathrm{St}$, allowing for multiple events of cluster shedding. This means that reducing $D$ with smaller $\sigma$ and lower $V_0$ would be preferable when $L$ is limited by the finite spatial size of the trapped condensate. We empirically confirmed it in our efforts to optimize the appearance probability of von K\'arm\'an vortex street. However, when $V_0$ was too close to $\mu$, the shedding pattern became excessively stochastic and $\kappa>1$ cluster emission was less likely~\cite{reeves2}. Longer streets of $\kappa=2$ vortex clusters were indeed observed with a smaller and harder obstacle [Fig. 4(a)], but its appearance probability was lower than 5\% and we noticed that the Gaussian beam profile was not clean. A larger condensate is definitely beneficial, but having $R/\xi \sim 10^3$ is experimentally challenging.

Second, because the velocity window for von K\'arm\'an street is quite narrow in $v/v_c$, precise control of the condensate motion is extremely important. Uncontrolled small dipole oscillations of the condensate can make observation of von K\'arm\'an street elusive. In our trap, dipole oscillations of $1~\mu$m corresponds to relative velocity oscillations of $\approx 0.07$~mm/s. For the same reason, it would be highly desirable to have a homogeneous sample, at least, along the obstacle trajectory. Assuming the universality of vortex shedding dynamics, one might consider dynamic control of $v$ and $V_0$ for constant $v/v_c$.

In conclusion, we have observed vortex cluster shedding in a Bose atomic superfluid and its transition from regular to turbulent shedding with increasing obstacle velocity. This work reveals the striking similarities between a superfluid and a classical fluid in vortex shedding dynamics. We expect that our work can be directly extended with even larger samples and various obstacle diameters to investigate the universality of the vortex shedding dynamics~\cite{reeves1}.

\begin{acknowledgements} 

We thank A.S.~Bradley and M.T.~Reeves for their valuable suggestions and discussion. This work was supported by IBS-R009-D1.

\end{acknowledgements}

\newpage

\widetext

\begin{center}
\textbf{\large Supplemental Material}
\end{center}

\setcounter{equation}{0}
\setcounter{figure}{0}
\setcounter{table}{0}
\setcounter{page}{1}
\makeatletter
\renewcommand{\theequation}{S\arabic{equation}}
\renewcommand{\thefigure}{S\arabic{figure}}

\begin{figure}[ht]
	\includegraphics[width=9.5 cm]{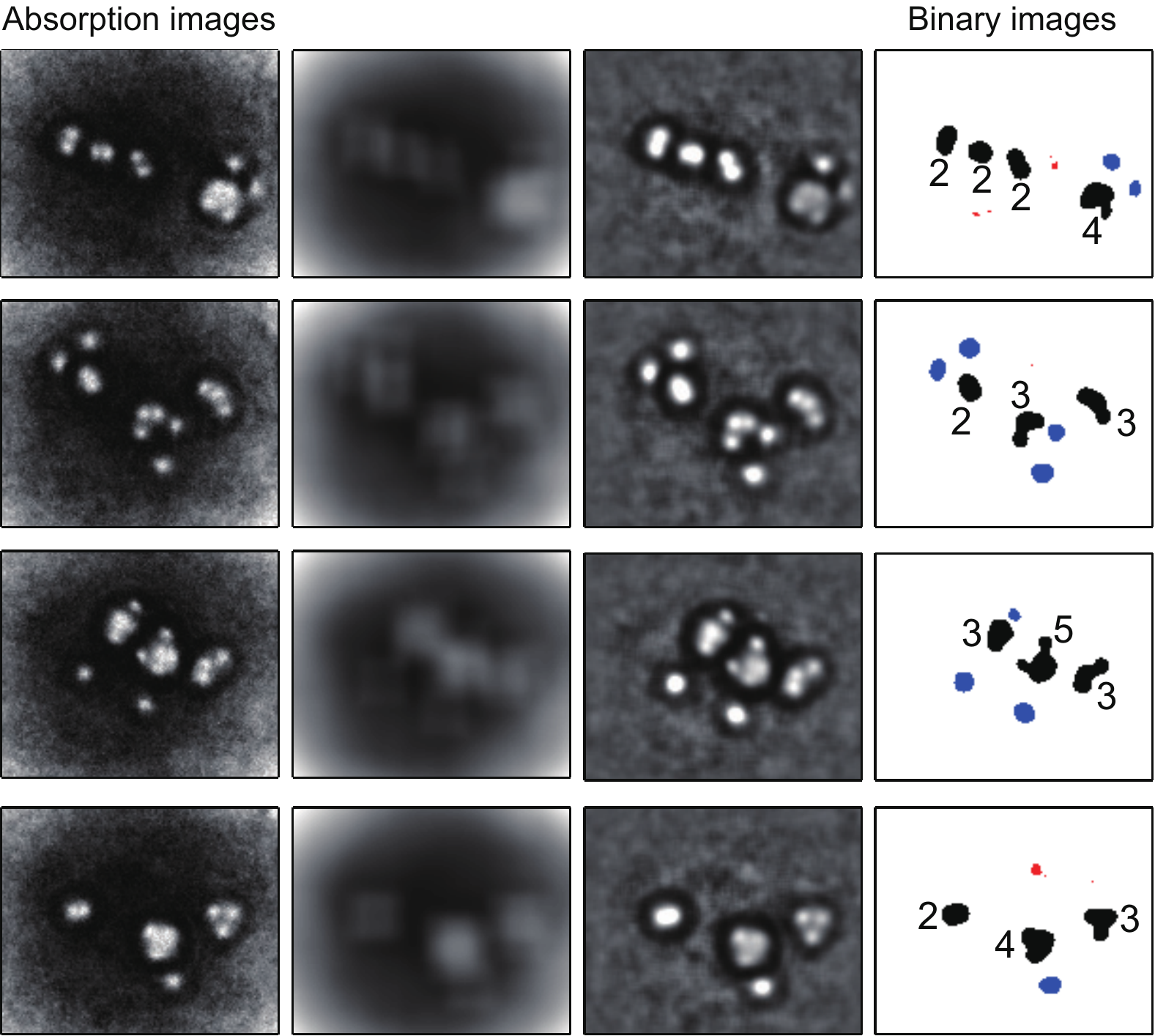}
	\caption{Determination of cluster charge. The first left column shows absorption images of condensates taken after 36~ms time of flight. In our imaging, the condensate radially expands by a factor of 2.3 and the FWHM of the density-depleted core of a singly charged vortex appears about 25~$\mu$m. The absorption images are divided by their boxcar-smoothened duplicates (second column) and the resultant images  (third column) are transformed into binary images for a certain threshold value, where the density-depleted regions appear as  particles of various sizes. The box width in the smoothening was set to be 60~$\mu$m, letting two nearby vortices merged into one particle when their separation is smaller than the core diameter of an individual vortex.  The numbers in the binary images denote the charge numbers of clusters.  Blue particles denote isolated single vortices. Red particles are imaging defects, ignored in our cluster counting.}
	\label{FigS2}
\end{figure}

\begin{figure}[ht]
	\includegraphics[width=11cm]{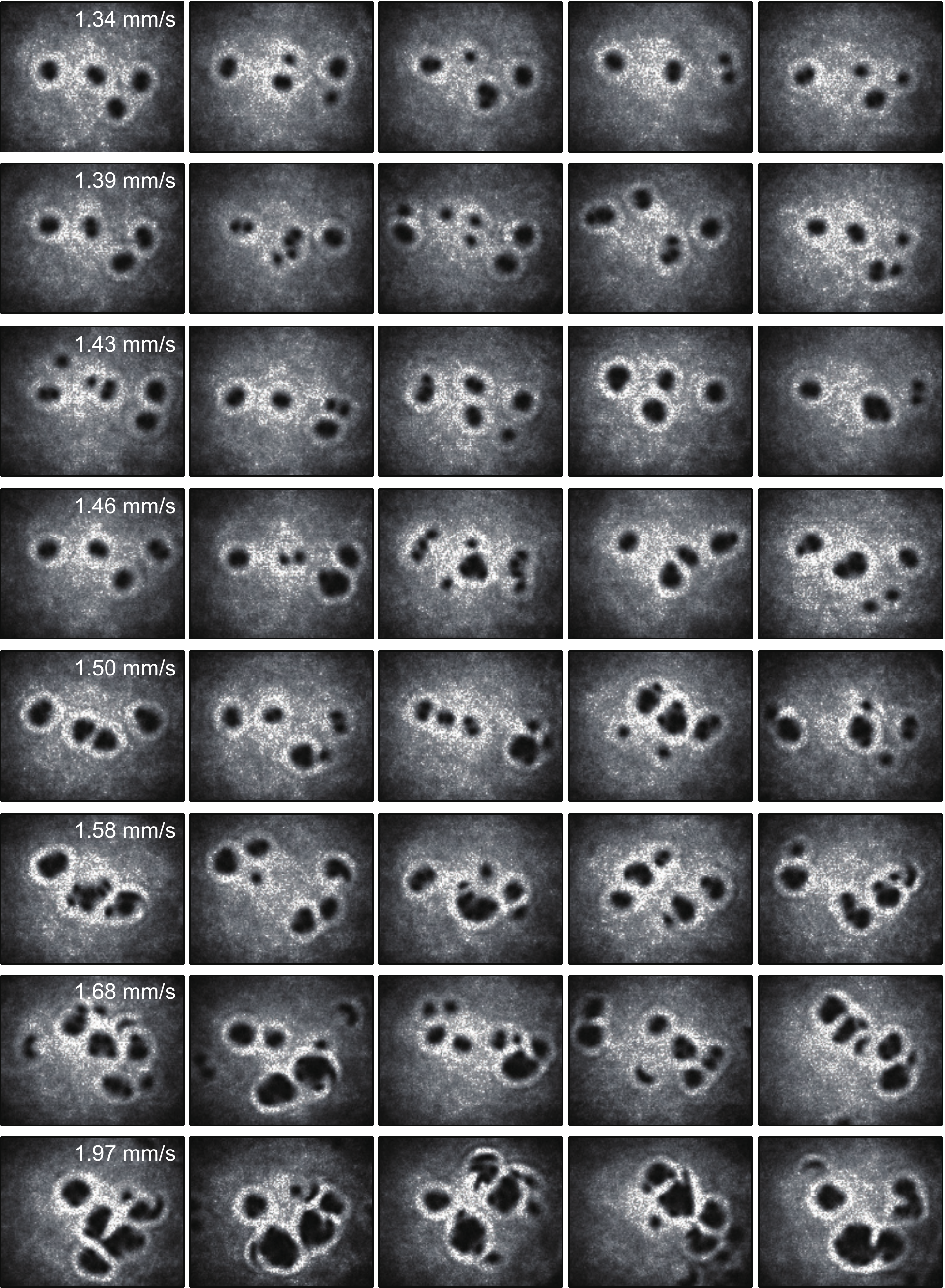}
	\caption{Additional image data of the shedding pattern for various $v$. Interestingly, a street pattern of four $\kappa$=3 clusters are observed in the first left image for $v=1.50$~mm/s, which happened very rarely in our experiment.}
	\label{FigS3}
\end{figure}

\begin{figure}[ht]
	\includegraphics[width=2.3cm]{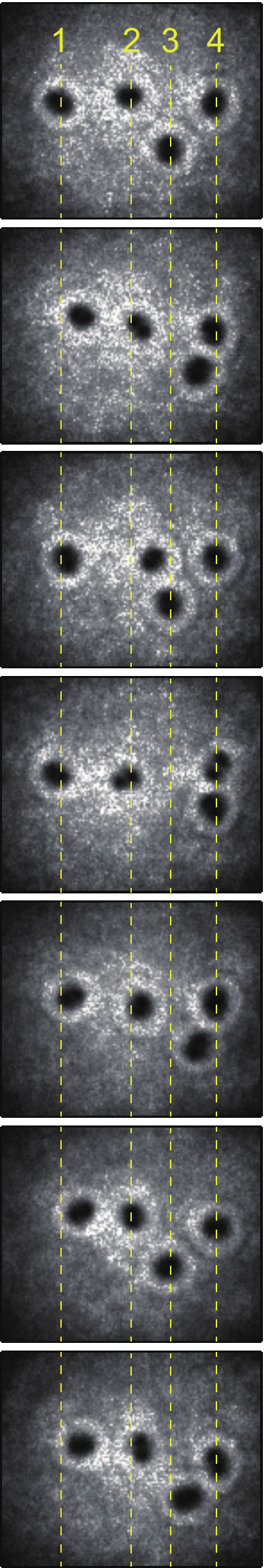}
	\caption{Images obtained at $v=1.39$~mm/s, showing a regular shedding pattern of four $\kappa$=2 vortex clusters. The vertical dashed lines indicate the horizontal positions of the vortex clusters in the top image. All of the last, i.e., most right clusters are located at the same horizontal position, indicating that they might have been released when the obstacle laser beam was turned off.}
	\label{FigS3}
\end{figure}


\begin{references}
\bibitem{tof} In order to improve the visibility of vortex cores, we took a two-step releasing procedure where the condensate expands first in the optical dipole trap for 12~ms by turning off the magnetic potential~\cite{kwon1}. In the imaging, the radial extent of the condensate increases by a factor of 2.3 and the full width at half maximum of the density-depleted hole of a singly charged vortex appears about 25~$\mu$m.
\end{references}
\end{document}